\documentclass[twocolumn,showpacs,preprintnumbers,amsmath,amssymb]{revtex4}
\usepackage{graphicx}
\usepackage{dcolumn}
\usepackage{bm}

\begin{document}

\title{Helix untwisting and bubble formation in circular DNA }

\author{ Marco Zoli }
\affiliation{
School of Science and Technology - CNISM \\  Universit\`{a} di Camerino, I-62032 Camerino, Italy \\ marco.zoli@unicam.it}

\date{\today}

\begin{abstract}
The base pair fluctuations and helix untwisting are examined for a circular molecule. A realistic mesoscopic model including twisting degrees of freedom and bending of the molecular axis is proposed. The computational method, based on path integral techniques, simulates a distribution of topoisomers with various twist numbers and finds the energetically most favorable molecular conformation as a function of temperature. The method can predict helical repeat, openings loci and bubble sizes for specific sequences in a broad temperature range. Some results are presented for a short DNA circle recently identified in mammalian cells.
\end{abstract}

\pacs{87.14.gk, 87.15.A-, 87.15.Zg, 05.10.-a}

\maketitle

\section*{I. Introduction}

The conformational properties of DNA and its biological functioning depend on key parameters as persistence length and helical repeat which, in turn, are sequence dependent and also vary with temperature and counterion concentrations \cite{gray,volo97}. Empirical approaches have been developed to quantify them in some cases. For circular molecules, common to mithocondrial DNA, bacterial plasmids and many viral genomes, measurements of cyclization probabilities and statistical analysis of topoisomers distributions \cite{shore1} remain the fundamental methods. Recently a new type of extra chromosomal circular (ecc) DNA consisting of short sequences, $ \sim 200 - 400 $ base pairs (\emph{bps}), has been identified in mouse and human cells \cite{dutta} remarkably suggesting that a large variability may occur in DNA of somatic tissues.
Topoisomers distributions are caused by thermal fluctuations which convert, by ligase, the nicked molecules to covalently closed circles each of them having a peculiar linking number ($Lk$) that defines its topological state \cite{bates}. Gel electrophoresis techniques have been used to estimate the average helix rotation angles in bacteriophage PM2 and \emph{E.coli} plasmid \cite{depew}. Later on, it has been shown that the double helix unwinds linearly with temperature up to the premelting regime \cite{duguet}.
Beyond being essential for DNA characterization, a precise knowledge of helical pitch and twist density is required to design nanostructures efficiently releasing anticancer drugs \cite{zhao} and molecules with functional properties \cite{metz07,kumar}.

These issues can be theoretically approached by mesoscopic models, treating DNA at the level of the fundamental \emph{bps} interactions, which may predict the energetically most convenient helical repeat ($h$) for a given sequence and ambient conditions. In this paper, I propose a  method based on the path integral formalism \cite{fehi} which can be applied to any circular molecule. The model, for $N$ \emph{bps}, includes \emph{both} twisting \emph{and} molecule bending to provide a realistic description of the effective interactions. This investigation extends a previous work \cite{i11} in which the thermodynamics of a short DNA had been computed, in a fixed plane representation, for a fixed value of the helical repeat. While we refer to that work for more details concerning the path integral method, here
the computation simulates a large topoisomers distribution of short circular DNAs with variable supercoiling degree and finds, at any  temperature, the most probable twisted geometry on the basis of a macroscopic constraint, the second law of thermodynamics. As the method calculates the thermal displacements of the \emph{bps} with respect to the ground state, we can also monitor the formation of fluctuational openings along the sequence. In particular, the location of the opening sites and the size of the bubbles can be determined at various temperatures.
These findings may be relevant also in view of the possibility to predict the sequence specific starting sites for biological functions such as transcription which require DNA unwinding and bubbles formation \cite{choi,erp2}.

The model is described in Section II while the results for a specific sequence of Ref.\cite{dutta} are discussed in Section III. Some final remarks are given in Section IV.

\section*{II. Model for Circular DNA}

In general, $Lk=\,Tw + Wr$, where $Tw$ is the twist accounting for the coiling of the individual strands around the helical axis and $Wr$ is the writhe measuring the spatial coiling of the axis itself \cite{ful1,marko1,mukamel}. While $Tw$ and $Wr$ (not necessarily integer numbers) refer to the molecular geometry, the integer $Lk$ is independent of the specific geometry. $Lk$ is given with respect to the \emph{relaxed linking number} of the least distorted topoisomer, $Lk_0 \approx \, N/h$. Thus, it is  $\sigma=\,(Lk - Lk_0)/Lk_0$  which measures the molecules supercoiling with almost all living beings keeping their DNA in a $\sigma < 0$ supercoiled state. This is essential to biological processes such as replication and transcription requiring the helix unwinding as a key step to favor the binding of proteins.

Short linear DNA, below $\sim 500$ \emph{bps}, shows a lower probability than long sequences to ligate into circles as a consequence of the intuitive fact that bending smaller fragments has a higher energy cost.
However, whenever ligation of the chain ends occurs in circles with decreasing diameters,
supercoiling increments are due to twisting rather than writhing \cite{shore1}. In fact, unwinding (or overwinding) the double helix requires essentially the same energy per unit length whereas the writhing involves crossings of the helix axis over itself and elastic deformations which are mostly confined in the apical loops. As the latter contain a larger proportion of the helix in smaller diameter circles, it is in shorter fragments that the energy required to change $Wr$ becomes increasingly higher than the energy associated to a change in $Tw$.
Hence, supercoiling increments are due to twisting rather than writhing.
Accordingly I assume that, for short circular DNA, the helix axis lies in a plane ($Wr=\,0$) while the degree of supercoiling measured by the linking number is attributed only to the twisting, $Lk \equiv Tw$.

\begin{figure}
\includegraphics[height=7.5cm,width=9.5cm,angle=-90]{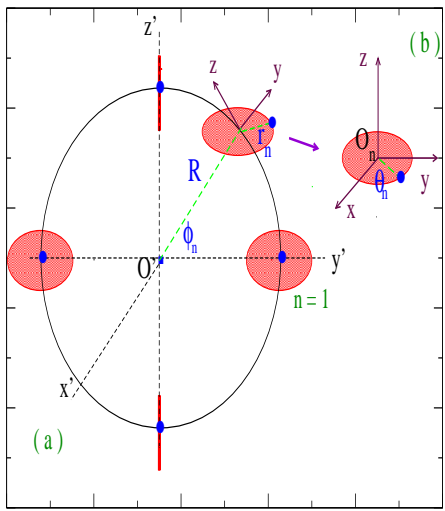}
\caption{\label{fig:1}(Color online) (a) Helicoidal model for circular DNA with bending planes. The blue filled circles are the pointlike \emph{bps} stacked along the molecule backbone. In the ground state all \emph{bps} lie on the circumference with ray $R$. The red-shaded areas are spanned by the fluctuational vectors whose amplitude is measured by $|r_n|$ for the $n-$ \emph{bp}.  $\phi_n$ is the bending of the $n-$ \emph{bp} plane with respect to the $(x',y')$ plane, $\textbf{x'}$ being normal to the sheet plane. (b) Local reference system for the $n-$ \emph{bp}. $\theta_n$ measures the twisting around the molecule backbone. The z-axis is tangent to the ground state circle.}
\end{figure}

Fig.~\ref{fig:1} displays the fragment with $N$ \emph{bps} whose equilibrium separations are regularly arranged in a circle with center point $O'$ and ray $R$ lying in the $(y',z')$ plane. The \emph{bp} fluctuation is described by an inter-strand vector displacement $\textbf{r}_n$  ($n$ numbers the \emph{bps} in real space)  measuring the pair mates separation with respect to the ground state. The latter is recovered when $|\textbf{r}_n|=\,0$ for any $n$.
In general, $\textbf{r}_n$ spans an orbit whose center $O_n$ stays on the ground state circle hence, $\overline{O'O_n}=\,R$.
With respect to $O'$, the fluctuation vector is:

\begin{eqnarray}
& &\textbf{t}_n =\, \Bigl( \bigl({t}_n \bigr)_{x'}, \bigl({t}_n\bigr)_{y'},  \bigl({t}_n\bigr)_{z'} \Bigr)
\, \nonumber
\\
& &\bigl({t}_n \bigr)_{x'} =\, |\textbf{r}_n| \cos\phi_n \cos\theta_n \, \nonumber
\\
& &\bigl({t}_n\bigr)_{y'} =\,(R + |\textbf{r}_n|\sin\theta_n) \cos\phi_n
\, \nonumber
\\
& &\bigl({t}_n\bigr)_{z'} =\,(R + |\textbf{r}_n|) \sin\phi_n \,.
\,
\label{eq:1}
\end{eqnarray}

The polar angle, $\theta_n =\, (n - 1) 2\pi {{Tw}/ N} + \theta_S$, measures the $n-$ \emph{bp} twisting around the molecule backbone with $h \,\equiv \,N/ Tw$.
$\theta_S$ is the twist of the first \emph{bp} along the stack.
As one turn of the helix hosts $h$ \emph{bps}, it matters to know which twist angle is associated to the starting sequence site: in fact the choice of the initial twist may affect the fluctuational amplitudes at the successive sites. Then, I integrate over a set of $\theta_S$ values, thus weighing an ensemble of distinct rotational conformations which contribute to the partition function.

The azimuthal angle, $\phi_n =\, (n-1){{2 \pi} / N}$, defines the (counterclockwise) rotation of the $n-th$ fluctuational orbit with respect to the $(x',y')$ plane. The latter contains the orbits of the fragment ends, the $n=\,1$ and $n=\,N+1$ \emph{bps}, which overlap due to the closure condition holding for the DNA ring.
Base pairs, which would be distant along the stack in a chain model, become closer in a circular model that accordingly accounts for stabilizing long range effects \cite{yera1}.

In general the shape of the fluctuational orbits is function of the bending $\phi_n$ and changes from circular ($\phi_1=\,0$) to a straight line ($\phi_{(N+1)/4}=\,\pi / 2$) being elliptic for $0 < \phi_n < \pi / 2$. In fact only a subset of orbital points do represent effective \emph{bps} fluctuations: for instance, the $n=\,1$ orbit is in principle circular but, being $\theta_n$ a discrete variable, only those points corresponding to the chosen set of $\theta_S$ values do correspond to real fluctuational states for the $n=\,1$ \emph{bp}.
With this caveat, one notices that the orbit size is determined by the fluctuation amplitude $|\textbf{r}_n|$ which is temperature dependent. While large thermal fluctuations may occur for any $n-$ site, the condition $|\textbf{r}_n| < R$  should be anyway fulfilled in the computation so as to ensure that an overall ring shape is preserved for the DNA fragment \cite{note1}.

Then, given a circular DNA sequence with $N$ \emph{bps}, the geometry is essentially determined by $R$ and $Tw$.
$R / N$ sets the \emph{bps} density along the molecule backbone.  Assuming $R=\,80 {\AA}$ the sequence length is consistently taken of order $50 nm$ which is also the persistence length of short DNA fragments recently measured at room temperature \cite{volo11}. $Tw / N$ sets the torsional stress of the double helix. Both parameters are incorporated in the $\textbf{t}_{n}$ variables.

Using the latter, I represent the system in Fig.~\ref{fig:1} by an extended Peyrard-Bishop (PB) Hamiltonian \cite{pey2}.
The model is treated by the finite temperature path integral method  \cite{i09} assuming that the \emph{bps} radial displacements are one dimensional paths $x(\tau_i)$ with the imaginary time $\tau_i \in [0, \beta]$, $\beta=\,(k_B T)^{-1}$, $k_B$ is the Boltzmann constant and $T$ is the temperature. The index $i$ numbers the \emph{bps} along the time axis. The periodic condition,  $x(\tau_i)=\, x(\tau_i + \beta)$, allows to Fourier expand the paths and accounts for the closure of the DNA sequence into a ring.
Partitioning the $\beta$ length in $N$ intervals, the space-time mapping of the fluctuational vectors in Eq.~(\ref{eq:1}) is performed through:

\begin{eqnarray}
& &\pm |\textbf{r}_n| \rightarrow \, x(\tau_i) ;  \, \, \, \pm |\textbf{r}_{n-1}| \rightarrow \, x(\tau_i - \beta /N)\, \nonumber
\\
& &\pm |\textbf{t}_n| \rightarrow \, \eta(\tau_i) ;  \, \, \, \pm |\textbf{t}_{n-1}| \rightarrow \, \eta(\tau_i - \beta /N).
\label{eq:5}
\end{eqnarray}

Hence the partition function of our system is written, in terms of the fluctuations amplitudes $\eta(\tau_i)$, as:

\begin{eqnarray}
& &Z =\,\oint \mathfrak{D}x \sum_{\theta_S} \exp\Biggl\{- \beta \sum_{i=\,1}^{N}  \Bigl[{\mu \over 2}\dot{\eta}_i^2  + V_M[\eta_i] + \,V_{sol}[\eta_i] \, \nonumber
\\
& & +  V_S[\eta_i,\eta'_i] \Bigr] \Biggr\}\,\, \nonumber
\\
& & V_M[\eta_i]=\,D_i \bigl[\exp(-a_i (\eta_i - R) - 1 \bigr]^2 \, \nonumber
\\
& & V_{sol}[\eta_i] =\, - D_i f_s \bigl[\tanh\bigl((\eta_i - R)/\l_s \bigr) - 1 \bigr] \, \nonumber
\\
& &V_S [\eta_i,\eta'_i]=\,K G_{i, i-1} \cdot \bigl(\eta_i - \eta'_i \bigr)^2 \, \nonumber
\\
& &G_{i, i-1}=\,1 + \rho_{i, i-1} \exp\bigl[-\alpha_{i, i-1}(\eta_i + \eta'_i - 2R)\bigr]
\, \nonumber
\\
& &\eta_i \equiv \, \eta(\tau_i) \, ; \, \eta'_i \equiv \, \eta(\tau_i - \beta /N) \, .
\label{eq:6}
\end{eqnarray}

The measure $\mathfrak{D}x$, which normalizes the kinetic action, is a multiple integral over the path Fourier coefficients \cite{i10}.
$\mu=\,300 amu$ is the reduced mass and $K=\,20meV {\AA}^{-2}$  is the harmonic force constant both for AT- and GC- \emph{bps} \cite{theo02}.
The Morse potential $V_M$ models the hydrogen bonds between complementary strands with site dependent effective pair dissociation energy $D_i$ and inverse length $a_i$.
Setting, $D_{AT}=\,30meV$ and $D_{GC}=\,45meV$, the hydrogen bond energies are above $k_BT$ at room temperature.
Further I take, $a_{AT}=\,2.4{\AA}^{-1}$ and $a_{GC}=\,2.7{\AA}^{-1}$.
The Morse plateau implies that, if all fluctuations are larger than  $a_i^{-1}$, the open strands can go in principle infinitely apart. As strands recombination may instead occur in solution, the solvent term  $V_{sol}$ is introduced \cite{druk,i11} to enhance the height ($f_s=\,0.1$)  and tune the width ($l_s=\,5{\AA}$) of the barrier for pair dissociation.

Adjacent \emph{bps} along the molecule stack interact via the potential $V_S$ which includes \emph{heterogeneity} in the anharmonic parameters  $\alpha_{i, i-1}$ and $\rho_{i, i-1}$. $\alpha_{i, i-1}/ a_i \ll 1$ ensures that the $V_S$ range  is larger than that of $V_M$.  Whenever one of the fluctuations is such that, $\eta_i - R > \alpha_{i, i-1}^{-1}$, the hydrogen bond breaks and the stacking coupling in Eq.~(\ref{eq:6}) drops from $\sim K(1 + \rho_{i, i-1})$ to $\sim K$. Then, also the next \emph{bp} tends to open as both pair mates are less closely packed along their respective strands. A reduced stacking implies a softer stretching frequency (smaller contribution to the free energy) hence, the cooperative formation of fluctuational openings along the backbone is measured by an entropic gain \cite{theo10}. As the latter is expected to be larger for AT \emph{bps}, heterogeneous stacking anharmonicity is
modeled by taking:
$\alpha_{AT, AT}=\,0.2{\AA}^{-1}$, $\rho_{AT, AT}=\,25$, $\alpha_{AT, GC}=\,0.3{\AA}^{-1}$, $\rho_{AT, GC}=\,15$, $\alpha_{GC, GC}=\,0.4{\AA}^{-1}$, $\rho_{GC, GC}=\,1$.

Somewhat different sets of parameters have be used in other studies to test the (homogeneous) anharmonic PB model albeit without rotational degrees of freedom \cite{campa,pey9,singh,handoko}. Morse parameters and harmonic force constants have been recently obtained also for RNA, fitting the PB model to the experimental melting temperatures \cite{weber}.

Computing Eq.~(\ref{eq:6}) amounts to sample the ensemble of molecule states where each state, defined by a set of Fourier coefficients, is a point in the path configuration space. About $2\cdot 10^6$ paths for every \emph{bp} are included in the computation at high $T$.

\section*{III. Results}

The model is applied to the micro DNA sequence with $N=\,184$ \emph{bps}  ($\sim 46 \%$ GC content) of  Ref.\cite{dutta}:

\begin{eqnarray}
& &AGGG{AA}GGGGGAGAAATC{AA}CTTTCCC \, \nonumber
\\ \,
& &ACA{AT}CCTACAACT{AT}T\overline{\textbf{C}}AAAAAGC{TT}\, \nonumber
\\ \,
& &AGTGGGAGG{TA}CAGGAGGTGG{AA}GCAC \, \nonumber
\\ \,
& &GGTGCC{T\overline{\textbf{T}}}CTTATCAC{AA}GCAGCTCT{TT}\, \nonumber
\\ \,
& &CGACAAGCCTC{TT}CGTGCTTCTC{TA}\overline{\textbf{A}}\, \nonumber
\\ \,
& &GC\overline{\textbf{T}}TTTTGA{AT}AGTTGTAGGA{TT}GTGG \, \nonumber
\\ \,
& &GAAAG{TT}GATTTCTCCCCC{TT}C
\label{eq:4}
\end{eqnarray}

Given the set of potential parameters the entropy, $S=\,k_B \beta^2 d [\beta^{-1} \ln Z] / d\beta$, is computed at a initial temperature  ($T=\,300K$) for a sufficiently large ensemble of DNA conformations. At the successive $T$ a new molecule ensemble is generated (consistently with the fact that the path fluctuations are $T-$dependent) and $S$ is re-evaluated. If $S$ does not grow versus $T$, a new partition of the path configuration space is performed until the selected DNA conformations fulfill the second law of thermodynamics throughout the whole considered range, $T \in [300,370]K$ with a $1K$ step. A large distribution of topoisomers is simulated  by treating the helical repeat as a free parameter to be determined at any $T$.
Hence the entropy profiles provide a criterion to select the energetically most favorable helicoidal geometry corresponding to the most stable DNA conformation for specific ambient conditions.
While this method produces a remarkable growth of the CPU time with respect to the previous work \cite{i11}, it also offers a more appropriate computational scheme to determine the equilibrium properties of the molecule.

\subsection*{A. Helical Repeat}

\begin{figure}
\includegraphics[height=7.5cm,width=9.5cm,angle=-90]{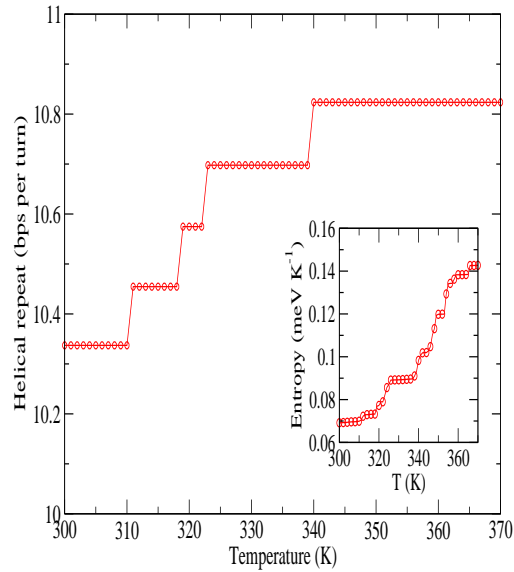}
\caption{\label{fig:2}(Color online) Helical repeat for the sequence in Eq.~(\ref{eq:4}). Inset: entropy versus temperature computed via Eq.~(\ref{eq:6}).}
\end{figure}

The helical repeat for the relaxed most probable topoisomer is first obtained at $T=\,300K$ and the change in $h$ is next computed at any larger $T$.
The theoretical approach simulates the experimental method by Depew and Wang \cite{depew}. The results are shown in Fig.~\ref{fig:2}. Remarkably the predicted $h$ values are in the range of those typical for DNA under physiological conditions  \cite{wang}. A stair-like pattern is found for $h$ with four incremental steps at $T_{1,2,3,4}=311, 319, 323, 340K$. The corresponding changes in the  helix twist angle $\bar{\theta}$ are: $\delta \bar{\theta} / \delta T_{1}=\,-0.032^\circ K^{-1}$, {\,} $\delta \bar{\theta} / \delta T_{2}=\,-0.043^\circ K^{-1}$, {\,} $\delta \bar{\theta} / \delta T_{3}=\,-0.078^\circ K^{-1}$, {\,} $\delta \bar{\theta} / \delta T_{4}=\,-0.022^\circ K^{-1}$ (per \emph{bp}), respectively. The average unwinding over the range $[300,\,340]K$  is $\delta \bar{\theta} / \delta T=\, -0.038^\circ K^{-1} bp^{-1}$ with the largest contribution arising at $T=\,323K$.
Previous studies \cite{duguet} have found a T-linear unwinding up to the pre-melting regime with $\delta \bar{\theta} / \delta T \sim \, -0.01 ^\circ K^{-1} bp^{-1}$, albeit for much longer sequences.
Note that, while our calculation determines the untwisting of short molecules, the cited experiments have only provided an average estimate of $h$ in sequences with thousands or more \emph{bps} for which closed and open segments may coexist below and inside the denaturation regime.

The entropic gains, see inset in Fig.~\ref{fig:2}, are responsible for the helix untwisting. Their values, calculated at the four temperature steps, are: $\Delta S(T_{1})=\,5 \cdot 10^{-4} meV K^{-1}$,  $\Delta S(T_{2})=\,3.6 \cdot 10^{-3} meV K^{-1}$,  $\Delta S(T_{3})=\,9.2 \cdot 10^{-3} meV K^{-1}$,  $\Delta S(T_{4})=\,5.1 \cdot 10^{-3} meV K^{-1}$. The smallness of the entropy increments suggests that helix denaturation is an overall smooth phenomenon in agreement with the conclusions of a recent neutron scattering analysis \cite{theo11}.

\subsection*{B. Thermal Bubbles}

Thermal fluctuations produce local openings which are similar to transcriptional bubbles starting at biologically active sites \cite{choi}. Location and size of denaturation bubbles depend on ambient conditions and sequence specificity \cite{kowalski,benham,krueg} with AT rich regions being more capable to release the torsional stress of supercoiled DNA \cite{zocchi3,metz10}. Bubble size distributions have also been used to determine the hydrogen bonds and stacking free energies of the \emph{bps} by stochastic optimization techniques \cite{metz11}.

\begin{figure}
\includegraphics[height=8.0cm,width=9.5cm,angle=-90]{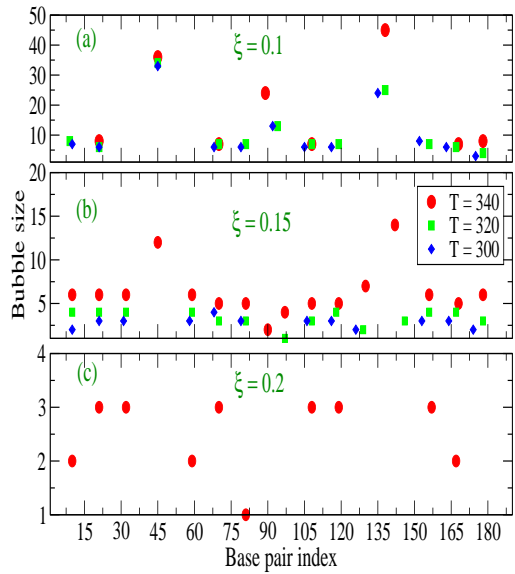}
\caption{\label{fig:3}(Color online) Number of consecutive base pairs whose stretchings are larger than  $\xi$ (in ${\AA}$) with respect to the ground state. The bubble sizes are $T$ dependent. The symbols abscissas mark the central base pair in the bubbles.}
\end{figure}

In the path integral method, position and growth of the bubbles can be monitored versus $T$ by computing the path fluctuations with respect to the ground state of the circular molecule. The path ensemble averaged \emph{bps} displacements $<\eta(\tau_i)>$ are compared to a threshold $\xi$: if,  $<\eta(\tau_i)> - R \geq \xi$, the $i-$\emph{bp} contributes to the bubble \cite{note2}. The $\xi$ values are arbitrary and may be tuned on the base of the model potential parameters for specific sequences \cite{i11a}. Thermal bubbles are formed, for the molecule in Eq.~(\ref{eq:4}), as a number of consecutive base pairs undergoes displacements which are of order of $\sim 0.1 - 0.2 {\AA}$. Larger amplitudes may be however expected for systems with higher percentages of AT base pairs. Fluctuational patterns are shown in Fig.~\ref{fig:3}. The bubble size is plotted versus the index marking the middle of the bubble itself. If the size is even, the abscissa indicates the \emph{closest to the middle} -\emph{bp} starting from the left in Eq.~(\ref{eq:4}). Fig.~\ref{fig:3}(a) shows that significant openings, larger than $\xi=\,0.1{\AA}$, exist already at room $T$ and are centered around the $i=45,\,135$ sites (overlined in Eq.~(\ref{eq:4})).
At $T=\,320K$ (and above), the  bubble centres are at $i=45,\,138$. At $T=\,300K$, the $i=\,45$ site hosts a CG-pair embedded in an extended $33$-\emph{bps} bubble with $64\%$ AT-pairs. The $i=\,135$ site hosts a AT-pair embedded in a $24$-\emph{bps} bubble with $67\%$ AT-pairs. The average AT content in the whole sequence is $56\%$. Hence, the main openings occur in the AT richest regions but GC pairs (inside those regions) cooperatively participate to the bubble formation.
The room $T$ fluctuations are smaller than $\xi=\,0.15{\AA}$ as both bubbles, centered at $i=45,\,135$,
disappear in Fig.~\ref{fig:3}(b). After heating the system at $T=\,340K$ the two bubbles show up again. However, the bubble sizes shrink with respect to the (a)-panel signalling that a path fluctuations subset is in the range $[0.1,\,0.15]{\AA}$.  Likewise, the bubble centered at $i=\,89$ contains $24$-\emph{bps} at $T=\,340K$ in the (a)-panel, whereas it spreads into much smaller openings in the (b)-panel at the same $T$. Only a few fluctuations are larger than $\xi=\,0.2{\AA}$ ((c)-panel) thus forming very localized bubbles  \cite{note3}. Altogether, our findings point to a substantial thermal stability of the short (ecc)-DNA with sizeable content of GC-pairs.

\section*{IV. Conclusion}

A realistic mesoscopic model incorporating twisting degrees of freedom and bending of the molecule axis has been developed  for circular DNA. Heterogeneous stacking anharmonicity stabilizes the molecule in the twisted geometry while permitting the formation of those local openings which release the torsional stress and sustain thermal fluctuations.
Path integral techniques have been developed to quantitatively predict the helix untwisting together with size and location of fluctuational bubbles. The base pairs fluctuational vectors are mapped onto time dependent paths contributing to the classical partition function. Our computation simulates a distribution of topoisomers with different twist numbers (as it is found in experiments) and finds the energetically most stable conformation as a function of temperature. The method has been applied to a small circular sequence with a relevant GC-content. While the bubbles are located in the AT-richest regions of the sequence, GC base pairs embedded in such regions can also experience sizeable fluctuations whose amplitude can be monitored at any temperature.  The predicted helical repeat presents a stair-like pattern whose incremental steps are associated to the entropic gains due to bubbles formation. While experimental information is becoming available to set accurate values for the effective model parameters, path integral computation is emerging as an efficient tool to investigate dynamics and stability conditions of DNA sequences.


\begin{references}
\bibitem{note2}
Bubble probability profiles can be obtained by path integral computation of $<\eta(\tau_i) - R>$.
\end{references}
\end{document}